%% file: main.tex
\documentclass[journal]{IEEEtran}

\usepackage{xcolor}
\usepackage[
    colorlinks=true,
    allcolors=blue
]{hyperref}

\usepackage{graphicx}
\usepackage{tikz}
\usetikzlibrary{shapes,arrows,arrows.meta,positioning,fit,backgrounds,calc,shadows.blur}

\definecolor{layer1}{HTML}{1F4E79}
\definecolor{layer1lt}{HTML}{D6E4F0}
\definecolor{layer2}{HTML}{2E75B6}
\definecolor{layer2lt}{HTML}{DEEBF7}
\definecolor{layer3}{HTML}{C55A11}
\definecolor{layer3lt}{HTML}{FCE4CC}
\definecolor{layer4}{HTML}{595959}
\definecolor{layer4lt}{HTML}{F2F2F2}
\definecolor{arrowcol}{HTML}{2E75B6}
\definecolor{goldarrow}{HTML}{C55A11}
\definecolor{protolbl}{HTML}{4472C4}

\usepackage{algorithm}
\usepackage{algpseudocode}

\usepackage{booktabs}
\usepackage{multirow}
\usepackage{array}
\usepackage{tabularx}

\usepackage{enumitem}

\usepackage{amsmath}
\usepackage{amssymb}
\usepackage[framemethod=default]{mdframed}
\newmdenv[backgroundcolor=blue!8,linecolor=gray!50,linewidth=0.8pt,roundcorner=3pt,innerleftmargin=8pt,innerrightmargin=8pt,innertopmargin=6pt,innerbottommargin=6pt]{chatuser}
\newmdenv[backgroundcolor=gray!8,linecolor=gray!50,linewidth=0.8pt,roundcorner=3pt,innerleftmargin=8pt,innerrightmargin=8pt,innertopmargin=6pt,innerbottommargin=6pt]{chatbot}

\usepackage{url}

\usepackage{subcaption}

\newcommand{\iotllm}{Grid-Orch}   
\newcommand{\mcp}{MCP}
\newcommand{\opendss}{OpenDSS}

\begin{document}



\title{Grid-Orch: An LLM-Powered Orchestrator for Distribution Grid Simulation and Analytics}

\author{\IEEEauthorblockN{Boming Liu, \emph{Member, IEEE}, Jin Dong, \IEEEmembership{Senior Member,~IEEE}, Jianming Lian, \emph{Senior Member, IEEE}} 
\IEEEpeerreviewmaketitle
\thanks{

This work was supported in part by the U.S. Department of Energy’s Office of Critical Minerals and Energy Innovation (CMEI) under Integrated Energy Systems Office Award No. 52746, and in part by the Office of Cybersecurity, Energy Security, and Emergency Response (CESER). Corresponding
Author: Jin Dong. B. Liu, J. Dong and J. Lian are with the Electrification and Energy Infrastructures Division, Oak Ridge National Laboratory, Oak Ridge, TN 37830 USA (e-mail: liub, dongj, lianj@ornl.gov). This manuscript has been authored by UT-Battelle, LLC, under contract DE-AC05-00OR22725 with the US Department of Energy (DOE). The US government retains and the publisher, by accepting the work for publication, acknowledges that the US government retains a non-exclusive, paid-up, irrevocable, worldwide license to publish or reproduce the submitted manuscript version of this work, or allow others to do so, for US government purposes. DOE will provide public access to these results of federally sponsored research in accordance with the DOE Public Access Plan.

}
 
}

\markboth{IEEE Open Access Journal of Power and Energy, 2026}
{Liu \MakeLowercase{\textit{et al.}}: Grid-Orch}

\maketitle

\begin{abstract}
The power distribution engineering workforce faces a projected shortage of up to 1.5 million engineers by 2030, creating urgent demand for more accessible analysis tools. This paper introduces \iotllm{}, a framework that bridges Large Language Models (LLMs) and power system simulation through the Model Context Protocol (\mcp{}), enabling engineers to perform complex distribution analyses via natural language. Using \opendss{} as the reference implementation, \iotllm{} provides 36 domain-specific tools across eleven categories---covering power flow, voltage analysis, quasi-static time-series (QSTS) simulation and automated optimization. A provider-agnostic LLM layer supports both cloud-hosted (Gemini, Claude) and locally deployed (Ollama, llama-cpp) models, enabling air-gapped operation for security-sensitive utility environments. Three optimization skills---capacitor placement, voltage violation analysis, and overvoltage mitigation---extend the platform beyond single-tool queries to multi-step engineering workflows. The developed \iotllm{} has an interactive web platform with chat-based interaction, a QSTS dashboard, and feeder topology visualization, with simulation results rendered inline. Workflow demonstrations show that distribution analyses formerly requiring hours of scripting, such as distributed energy resource (DER) interconnection screening,  complete in under two minutes through natural language, producing numerically identical results to direct \opendss{} scripting.\end{abstract}



\input{sections/introduction}
\input{sections/related_work}

\input{sections/architecture}
\input{sections/tool_library}
\input{sections/platform}
\input{sections/workflows}
\input{sections/discussion}

\vspace{-1em}
\section*{Acknowledgment}
\vspace{-0.5em}
The authors greatly acknowledge Yi Yang U.S. Department of Energy’s Office of Critical Minerals and Energy Innovation (CMEI), for her contributions to the conceptualization and review of this manuscript.

\bibliographystyle{IEEEtran}
\vspace{-1em}
\bibliography{references}

\end{document}

%% file: sections/introduction.tex
\section{Introduction}
\label{sec:introduction}

\IEEEPARstart{T}{he} power distribution engineering workforce faces an acute capacity crisis.
IEEE PES and Kearney project a shortage of up to 1.5~million qualified engineers by
2030~\cite{ieeekearney2024}, driven by accelerating grid modernization, mass retirements, and
explosive DER integration~\cite{useer2024,doe2024gridmod}.
The IEA reports that more than 60\% of energy companies cite labor shortages as a primary
operational constraint~\cite{iea2025employment}.
This imbalance (growing grid complexity against a shrinking workforce) calls for tools that
let engineers run distribution analyses without specialized scripting skills.

Existing simulation platforms fall short on accessibility.
\opendss{}~\cite{opendss2023}, the open-source distribution power-flow and
quasi-static time-series (QSTS) simulation tool, requires practitioners to author DSS scripts or
navigate its Python bindings.
Commercial alternatives (Synergi, CYME, PSCAD) offer graphical interfaces but carry substantial licensing costs that may be prohibitive for smaller utilities and research groups.
Neither category provides a conversational interface: engineers must still formulate correct API
calls rather than asking questions in natural language.

Recent advances in large language models (LLMs) demonstrate that natural language can serve as a
practical interface for technical software.
Function-calling capabilities~\cite{openai2023functions,yao2023react,wang2024agents} allow LLMs to reason over a tool
library and invoke domain-specific operations on behalf of the user.
Early power-system applications confirm the potential: LLM-based frameworks have been applied to
simulation orchestration~\cite{jia2025multiagent}, optimal dispatch~\cite{li2025optdispro}, grid
control~\cite{zhang2025gridagent,wen2025xgridagent}, and voltage regulation~\cite{jena2025llmvoltage},
establishing that LLMs can parse engineering intent, select tools, and interpret simulation outputs.

Despite this progress, four critical gaps remain unaddressed in the literature.

\begin{enumerate}[label=(\roman*)]

\item \textbf{No standardized tool-integration protocol.}
Existing integrations define bespoke function registries tied to a single codebase; no published
work adopts a vendor-neutral, community-standardized protocol for exposing simulator capabilities.

\item \textbf{Vendor lock-in.}
Current platforms couple their LLM layer to a specific proprietary API, precluding open-weight
or locally hosted models and creating dependency on external service pricing and availability.

\item \textbf{No comprehensive distribution analysis interface.}
Existing tools address isolated tasks; none provides a single interface covering load-shape
management, QSTS simulation, equipment sizing, topology visualization, and optimization.

\item \textbf{No support for air-gapped deployments.}
Utility OT networks routinely prohibit outbound cloud traffic; no existing LLM simulation tool
supports fully local operation with both inference and simulation running on-premises.

\end{enumerate}

\begin{figure*}[hbt]
  \centering
  \includegraphics[width=0.9\textwidth]{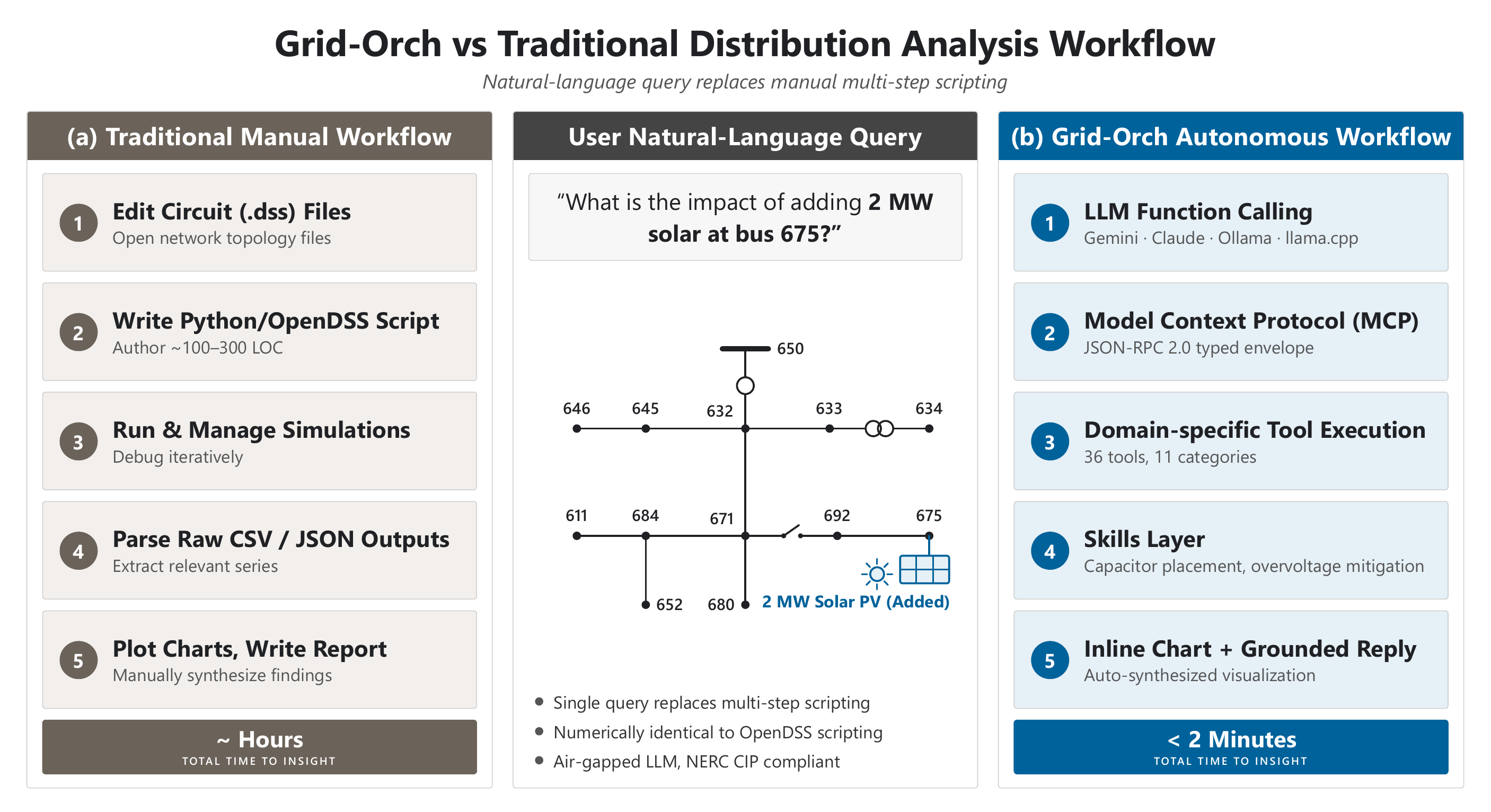}
  \caption{Comparative workflow for power distribution analysis. (a)~Traditional five-step manual process spanning several hours. (b)~\iotllm{} autonomous workflow completing the same analysis in under two minutes through five stages: LLM tool selection, \mcp{} routing, \opendss{} execution, Skills optimization, and synthesized visualization.}
  \label{fig:workflow_comparison}
\end{figure*}

The Model Context Protocol (\mcp{})~\cite{anthropic2024mcp}, an open standard for exposing tools to LLMs, addresses gap~(i) directly.
Because \mcp{} is model- and transport-agnostic, a single server can be queried by cloud or local
models without modification, resolving gap~(ii).
Building \iotllm{} on \mcp{} cleanly separates the
simulation layer (\opendss{}) from the inference layer (any compliant LLM).

This paper makes four primary contributions:

\begin{enumerate}

\item \textbf{\mcp{} adaptation for distribution simulation.}
We present an integrated adaptation of \mcp{} to a power system simulation engine, exposing 36
domain-specific tools across eleven categories through a standards-compliant server backed by \opendss{}.

\item \textbf{Multi-provider LLM layer with local-deployment support.}
\iotllm{} supports four backends (Gemini, Claude, Ollama, llama-cpp) via a provider-agnostic
abstraction, enabling fully air-gapped operation with no cloud dependency.

\item \textbf{Interactive web platform with integrated visualization.}
The web application delivers a chat interface, a QSTS results dashboard, and an interactive feeder topology map, all rendered inline during conversation.

\item \textbf{Skills framework for multi-step optimization workflows.}
A composable skills layer encapsulates voltage violation analysis, overvoltage mitigation, and capacitor placement as natural-language-invocable multi-tool pipelines. 
These skills reason over user prompts, decide which tools to call and in what order, and execute multi-step optimization workflows. Because the underlying OpenDSS engine supports three-phase unbalanced distribution systems, all tools and skills operate on full three-phase models; this enables accurate and trustworthy analysis of feeders such as the IEEE 13-bus and 123-bus test systems.

\end{enumerate}

Fig.~\ref{fig:workflow_comparison} illustrates the resulting workflow transformation: a five-step manual process spanning several hours is compressed into a single natural-language query completed in under two minutes.

Section~\ref{sec:related_work} reviews related work; Section~\ref{sec:architecture} describes the
system architecture; Section~\ref{sec:tool_library} details the tool library and
skills framework; Section~\ref{sec:platform} covers the web platform and
visualization; Section~\ref{sec:workflows} presents workflow demonstrations; and Section~\ref{sec:discussion} concludes with deployment considerations and future work.

%% file: sections/related_work.tex
\vspace{-1em}
\section{Related Work}
\label{sec:related_work}

\begin{table*}[!hbt]
  \caption{Comparison of \iotllm{} with Representative Related Systems on Distribution Analysis Dimensions. Systems differ in primary scope; see text for discussion.}
  \label{tab:comparison}
  \centering
  \renewcommand{\arraystretch}{1.25}
  \begin{tabular}{lllccc}
    \toprule
    \textbf{System} &
    \textbf{Focus} &
    \textbf{Protocol} &
    \textbf{Tools} &
    \textbf{Multi-LLM} &
    \textbf{Local Deploy} \\
    \midrule
    Grid-Agent~\cite{zhang2025gridagent}
      & Grid control          & Custom API     & $<$10   & No  & No    \\
    X-GridAgent~\cite{wen2025xgridagent}
      & Grid analysis         & Custom API     & $<$10   & No  & No    \\
    Jia et al.~\cite{jia2025multiagent}
      & Power flow sim.       & Custom API     & $<$5    & No  & No    \\
    OptDisPro~\cite{li2025optdispro}
      & Optimal dispatch      & Custom API     & $<$5    & No  & No    \\
    ChatGrid~\cite{jin2024chatgrid}
      & Grid visualization    & Custom API     & $<$5    & No  & No    \\
    Jena et al.~\cite{jena2025llmvoltage}
      & Voltage regulation    & Custom API     & $<$5    & No  & No    \\
    GridMind~\cite{jin2025gridmind}
      & ACOPF + N-1 contingency & PydanticAI   & 7       & Yes & No    \\
    \midrule
    \textbf{\iotllm{} (this work)}
      & \textbf{Full distrib.\ analysis} & \textbf{\mcp{}} & \textbf{36} &
      \textbf{Yes} & \textbf{Yes}  \\
    \bottomrule
  \end{tabular}
\end{table*}

\begin{figure*}[hbt]
  \centering
  \includegraphics[width=0.95\textwidth]{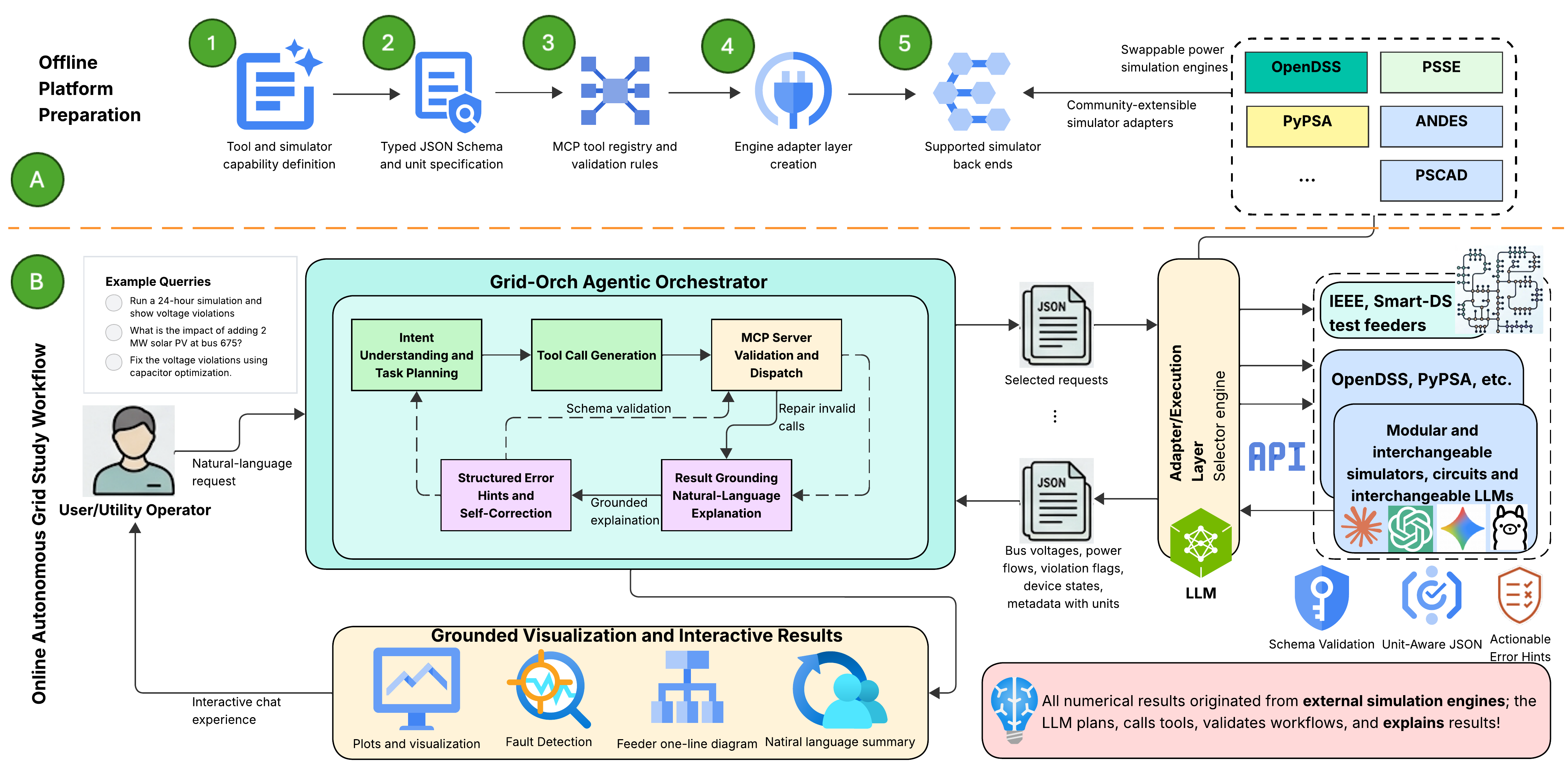}
  \caption{Overall Schematic of \iotllm{}, which translates natural-language grid-analysis requests into typed tool calls, validates them through a shared \mcp{} interface, dispatches them to swappable simulation engines, and grounds all responses and visualizations in structured simulator outputs.}
  \label{fig:scheme}
\end{figure*}

\subsection{LLMs in Power Systems}

The application of LLMs to power systems has grown rapidly, spanning control, planning, fault
diagnosis, and operator assistance~\cite{sarwar2025llmpower,chen2025pnnl}.
Agent-level frameworks~\cite{wang2024agents} underlie recent work on real-time grid
control~\cite{zhang2025gridagent,wen2025xgridagent}, simulation orchestration and
dispatch~\cite{jia2025multiagent,li2025optdispro}, ACOPF and contingency analysis~\cite{jin2025gridmind}, topology reasoning~\cite{bernier2025powergraph},
voltage regulation~\cite{jena2025llmvoltage}, control decision support~\cite{choi2024egridgpt}, and grid-state visualization~\cite{jin2024chatgrid}.
Across this body of work, integrations are task-specific or coupled to proprietary APIs; none
exposes a standards-compliant, auditable tool protocol to the LLM.

\vspace{-1em}
\subsection{The Model Context Protocol}

\mcp{}~\cite{anthropic2024mcp} standardizes LLM-to-tool communication via a capability-negotiation handshake, with demonstrated generalization to IoT
contexts~\cite{yang2025iotmcp} and analyzed security properties~\cite{mcp2025security}. \mcp{} is structured around three core primitives---\emph{tools}, \emph{resources}, and \emph{prompts}---which collectively enable standardized interaction between LLM agents and external systems.
General-purpose LLM orchestration frameworks such as LangChain~\cite{langchain2023} provide tool-calling abstractions but lack domain-specific schema validation and the standardized capability negotiation that \mcp{} offers.
During the preparation of this manuscript, the PowerAgent
project~\cite{zhang2025poweragent} released an open-source collection of
\mcp{} servers spanning multiple power system simulators, and has since grown
into a broader ecosystem adding a skill library, example agentic workflows, and agent-evaluation benchmarks. We have therefore added Table~\ref{tab:poweragent} to position \iotllm{} relative to this newly released platform across four concrete implementation dimensions.  

The comparison highlights complementary but distinct design objectives. PowerAgent~\cite{zhang2025poweragent} emphasizes breadth, supporting a wide range of transmission, dynamic, and electromagnetic-transient simulation platforms, including PSS/E, PowerWorld, PowerFactory, and PSCAD, while also pursuing power-system-oriented foundation-model development. In contrast, Grid-Orch was developed as an end-to-end agentic platform specifically for distribution-system analysis and operation. Its distinguishing contributions include native in-process optimization capabilities, tight integration of simulation, reasoning, and control workflows within a unified end-user web environment, and fully local inference as a first-class architectural requirement for privacy-sensitive and NERC CIP-constrained deployments. Accordingly, Grid-Orch extends beyond simulator connectivity to provide a complete, deployable workflow layer for autonomous distribution-system analysis, decision support, and optimization. Future work will further enhance interoperability by integrating selected PowerAgent \mcp{} tools into the \iotllm{} orchestration framework.

\begin{table}[!hbt]
  \caption{Qualitative Comparison of \iotllm{} and PowerAgent.}
  \label{tab:poweragent}
  \centering
  \renewcommand{\arraystretch}{1.2}
  \footnotesize
  \setlength{\tabcolsep}{4pt}
  \begin{tabularx}{\columnwidth}{>{\raggedright\arraybackslash}p{1.55cm}>{\raggedright\arraybackslash}X>{\raggedright\arraybackslash}X}
    \toprule
    \textbf{Dimension} & \textbf{\iotllm{}} & \textbf{PowerAgent} \\
    \midrule
    Simulator  &
      \opendss{} &
      13+ Engines \\
    Workflow orchestration &
      Native in-process skills (capacitor placement, voltage analysis, overvoltage mitigation) &
      Two example workflows; execution deferred to external frameworks (LangGraph, AutoGen, CrewAI) \\
    User interface &
      Integrated web platform: chat, result dashboard, topology map, exports &
      None shipped; integrates with third-party LLM clients \\
    Air-gapped support &
      Full local operation (Ollama, llama-cpp) as a design goal for NERC~CIP environments &
      Locally hosted models demonstrated in benchmarks; isolated deployment not documented \\
    \bottomrule
  \end{tabularx}
\end{table}

To the best of our knowledge, no existing open-source platform integrates a standardized tool protocol, multi-provider LLM support with air-gapped inference, a 36-tool distribution-system library, and a multi-step optimization Skills framework within a single system; Table~\ref{tab:comparison} summarizes the comparison with related work.
\vspace{-1em}

%% file: sections/architecture.tex
\section{System Architecture}
\label{sec:architecture}

\subsection{Overview}


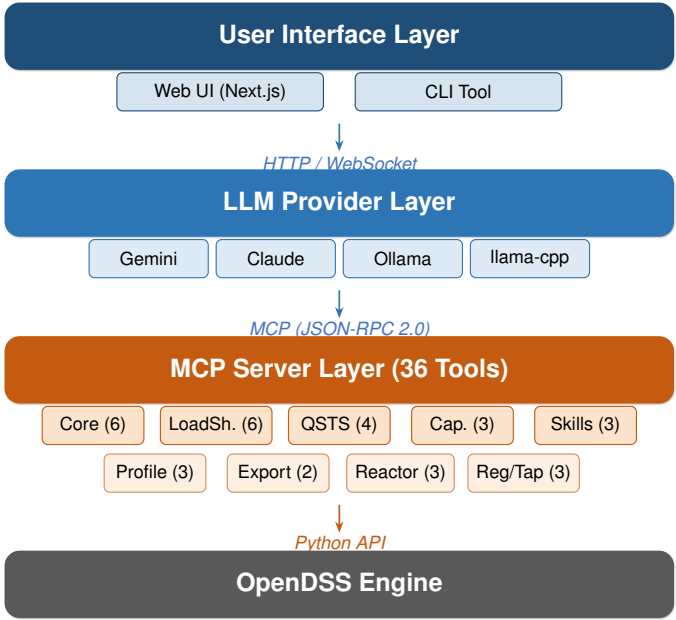
\begin{figure}[hbt]
  \centering
  \resizebox{\columnwidth}{!}{\input{tikz/fig_architecture}}
  \caption{\iotllm{} four-layer architecture. The \mcp{} boundary (dashed)
    decouples the LLM Abstraction Layer from the Simulation Engine Layer,
    allowing either side to be replaced independently.}
  \label{fig:architecture}
\end{figure}

Fig.~\ref{fig:scheme} illustrates the overall \iotllm{} workflow, in which natural-language requests are translated into validated \mcp{} tool calls, dispatched to swappable simulation engines, and returned as simulator-grounded responses and visualizations based on structured simulator outputs. Fig.~\ref{fig:architecture} further shows that this workflow is implemented through a four-layer architecture that cleanly separates user interaction from simulation.

\textbf{Layer~1 -- UI Layer.}
A React/Next.js application provides a persistent chat panel and a QSTS
results dashboard with voltage time-series plots, load-shape viewers, and a
force-directed feeder topology map.  The UI communicates exclusively with the
backend REST API, with no direct knowledge of the LLM provider or \opendss{}.

\textbf{Layer~2 -- LLM Abstraction Layer.}
A FastAPI backend hosts the provider-agnostic LLM service, running a
\emph{tool-use loop}: the LLM emits tool-call requests, the service dispatches
them to the \mcp{} layer, appends results as tool-response messages, and
repeats until the LLM produces a final text reply.  PostgreSQL persists chat
sessions and circuit metadata; MinIO holds binary circuit files.

\textbf{Layer~3 -- \mcp{} Server Layer.}
A dedicated Python process exposes all 36 domain tools through the Model
Context Protocol, owning JSON Schema validation, parameter coercion, and error
formatting.  It maintains a single shared \opendss{} manager instance so that
successive tool calls within one conversation operate on the same live circuit
state.

\textbf{Layer~4 -- Simulation Engine Layer.}
\opendss{}~\cite{opendss2023} performs all numerical computation via the
\emph{opendssdirect.py} bindings~\cite{opendss2023direct}, enabling
in-process deployment on Linux-based Docker containers.

The \mcp{} boundary acts as a \emph{capability firewall}: the LLM interacts
with the simulator only through the 36 defined tools, never through arbitrary
code execution.

\vspace{-1em}
\subsection{MCP Integration}
\label{subsec:mcp}

The Model Context Protocol is an open standard defining a JSON-RPC 2.0
transport for connecting LLMs to external tools and data
\cite{anthropic2024mcp}. Conceptually, MCP serves as a universal adapter between LLMs and external tools, analogous to how SCADA protocols standardize communication between control centers and field devices in power system operations.  \iotllm{} uses all three \mcp{} primitives.  The 36 simulation capabilities are registered as Tools with typed JSON Schema parameters.  Active circuit DSS files are exposed as Resources for topology queries.  A power-engineering Prompt template injects domain knowledge such as per-unit notation and standard voltage limits into every session.

The end-to-end interaction sequence is detailed in Fig.~\ref{fig:e2e_pipeline} (Section~\ref{sec:platform}). The server validates each incoming request against its JSON Schema, executes the corresponding handler, and returns a structured result to the LLM.  Multiple tool calls may be chained within a single user turn.

Error handling supports LLM self-correction: failed calls return structured hints enabling the model to recover without user intervention~\cite{qin2024toolllm} (detailed in Section~\ref{sec:tool_library}).

Beyond the vendor-neutrality and multi-provider support that motivated this work (Section~\ref{sec:introduction}), \mcp{} provides a standardized resource primitive for exposing non-tool data (e.g., circuit files) without custom endpoints~\cite{openai2023functions,langchain2023}.

\vspace{-1em}
\subsection{Multi-LLM Provider Support}
\label{subsec:providers}

The LLM Abstraction Layer implements a \emph{provider adapter} interface with
four concrete backends: \emph{Gemini (Google DeepMind), Claude (Anthropic),
Ollama (local Docker inference), and llama-cpp (GGUF-quantized models for
resource-constrained hosts)}.  All four adapters share identical upstream and
downstream interfaces---receiving an \mcp{} tool-schema list and message
history, returning either a final text reply or a list of tool calls---so
adding a new provider requires no changes to the \mcp{} layer or UI.  Utility
environments subject to NERC CIP cybersecurity standards \cite{nerc_cip_007} can select the Ollama
or llama-cpp adapter for fully \textbf{air-gapped operation}, ensuring circuit files and
simulation results never leave the security perimeter.

\vspace{-1em}
\subsection{OpenDSS Integration}
\label{subsec:opendss}

\iotllm{} accesses \opendss{}~\cite{opendss2023,dugan2011opendss} exclusively through the
\emph{opendssdirect.py} Python package~\cite{opendss2023direct}, which wraps
the native engine as an in-process shared library, eliminating COM interop
latency and enabling Linux-based Docker deployment.  The
\emph{OpenDSSManager} class maintains a single engine instance per server
process across circuit loading, power-flow solve, QSTS time-stepping, and
equipment state updates (capacitor switching, reactor placement, regulator tap
writes).  Security controls at the circuit-loading boundary validate that
requested file paths resolve to a whitelisted directory and block symbolic-link traversal, preventing path-traversal
attacks in multi-tenant deployments.

The current implementation maintains a single \opendss{} engine instance per server process; concurrent sessions are serialized through a request queue.  Multi-user deployments requiring parallel simulations can scale horizontally by running multiple backend containers behind a load balancer, each owning an independent engine instance.

%% file: tikz/fig_architecture.tex
\begin{tikzpicture}[
  >=Stealth,
  layer/.style={
    rectangle, draw, rounded corners=5pt, thick,
    minimum width=8.4cm, minimum height=0.82cm,
    font=\small\bfseries\sffamily, align=center, text=white,
    blur shadow={shadow xshift=0.3pt, shadow yshift=-0.5pt,
                 shadow blur steps=4, shadow blur radius=0.8pt}
  },
  comp/.style={
    rectangle, draw, rounded corners=2pt,
    minimum height=0.48cm, inner sep=3pt,
    font=\fontsize{6.5}{7.5}\selectfont\sffamily, align=center
  },
  proto/.style={
    font=\fontsize{6.5}{7.5}\selectfont\sffamily\itshape,
    text=protolbl
  },
  arr/.style={-{Stealth[length=4pt,width=3pt]}, semithick, color=arrowcol},
  garr/.style={-{Stealth[length=4pt,width=3pt]}, semithick, color=goldarrow},
]

\node[layer, fill=layer1, draw=layer1] at (0,0) (ui) {User Interface Layer};

\node[comp, fill=layer1lt, draw=layer1, minimum width=2.6cm]
      at (-1.5,-0.7) (webui) {Web UI (Next.js)};
\node[comp, fill=layer1lt, draw=layer1, minimum width=2.6cm]
      at (1.5,-0.7) (cli) {CLI Tool};

\draw[arr] (0,-1.1) -- (0,-1.45);
\node[proto] at (0,-1.6) {HTTP / WebSocket};

\node[layer, fill=layer2, draw=layer2] at (0,-2.1) (llm) {LLM Provider Layer};

\node[comp, fill=layer2lt, draw=layer2, minimum width=1.5cm] at (-2.4,-2.8) {Gemini};
\node[comp, fill=layer2lt, draw=layer2, minimum width=1.5cm] at (-0.8,-2.8) {Claude};
\node[comp, fill=layer2lt, draw=layer2, minimum width=1.5cm] at (0.8,-2.8) {Ollama};
\node[comp, fill=layer2lt, draw=layer2, minimum width=1.5cm] at (2.4,-2.8) {llama-cpp};

\draw[arr] (0,-3.2) -- (0,-3.55);
\node[proto] at (0,-3.7) {MCP (JSON-RPC 2.0)};

\node[layer, fill=layer3, draw=layer3] at (0,-4.2) (mcp) {MCP Server Layer (36 Tools)};

\node[comp, fill=layer3lt, draw=layer3, minimum width=1.28cm] at (-3.1,-4.9) {Core (6)};
\node[comp, fill=layer3lt, draw=layer3, minimum width=1.28cm] at (-1.55,-4.9) {LoadSh.\ (6)};
\node[comp, fill=layer3lt, draw=layer3, minimum width=1.28cm] at (0,-4.9) {QSTS (4)};
\node[comp, fill=layer3lt, draw=layer3, minimum width=1.28cm] at (1.55,-4.9) {Cap.\ (3)};
\node[comp, fill=layer3lt, draw=layer3, minimum width=1.28cm] at (3.1,-4.9) {Skills (3)};

\node[comp, fill=layer3lt!55, draw=layer3!70, minimum width=1.28cm] at (-2.32,-5.5) {Profile (3)};
\node[comp, fill=layer3lt!55, draw=layer3!70, minimum width=1.28cm] at (-0.77,-5.5) {Export (2)};
\node[comp, fill=layer3lt!55, draw=layer3!70, minimum width=1.28cm] at (0.77,-5.5) {Reactor (3)};
\node[comp, fill=layer3lt!55, draw=layer3!70, minimum width=1.28cm] at (2.32,-5.5) {Reg/Tap (3)};

\draw[garr] (0,-5.9) -- (0,-6.25);
\node[proto, text=goldarrow] at (0,-6.4) {Python API};

\node[layer, fill=layer4, draw=layer4] at (0,-6.9) (dss) {OpenDSS Engine};

\end{tikzpicture}%

%% file: sections/tool_library.tex
\vspace{-1em}
\section{Tool Library and Skills}
\label{sec:tool_library}
\label{sec:skills}

In the context of LLM--simulator integration, a \emph{tool} is a single callable operation, such as solving a power flow or reading a bus voltage, while a \emph{skill} is a multi-step workflow that chains several tools together to accomplish a complex engineering task (e.g., optimizing capacitor placement). Tools and skills are critical for OpenDSS integration because they provide a structured, auditable interface between the LLM's natural language understanding and the simulation engine's numerical computation: every analysis step is traceable to a specific tool invocation with typed inputs and outputs, thereby reducing the risk that the LLM fabricates numerical results.

\iotllm{} exposes 36 tools organized into 11 categories (Table~\ref{tab:tools}), collectively covering the distribution-engineering workflow from circuit loading through optimization. The LLM autonomously selects the appropriate tools from the available schema definitions.

\vspace{-1em}
\subsection{Tools}

\begin{table}[t]
  \caption{\iotllm{} \mcp{} Tool Library (36 tools, 11 categories)}
  \label{tab:tools}
  \renewcommand{\arraystretch}{1.2}
  \footnotesize
  \begin{tabular}{@{}lcp{4.8cm}@{}}
    \toprule
    \textbf{Category} & \textbf{Tools} & \textbf{Key Capabilities} \\
    \midrule
    Core Circuit      & 6 & Circuit loading, Newton--Raphson power flow, single-bus and system-wide voltage queries, circuit metadata \\
    LoadShape         & 6 & Create, edit, delete, and assign time-series load profiles to load objects \\
    QSTS Simulation   & 4 & Run quasi-static time-series simulations; retrieve voltage and loss time-indexed results \\
    Profile Library   & 3 & Browse and load built-in, NREL, and custom CSV load profiles \\
    Results Export    & 2 & CSV/JSON data export and HTML report generation \\
    Capacitor Mgmt   & 3 & Add/remove shunt capacitor banks at any bus \\
    Reactor Mgmt     & 3 & Add/remove shunt reactors for overvoltage absorption \\
    Regulator/Tap    & 3 & Read and adjust voltage regulator tap positions \\
    Circuit Library  & 2 & Load pre-packaged IEEE and SmartDS test feeders \\
    Topology         & 1 & Bus coordinates and branch connectivity for visualization \\
    Skill Invocation & 3 & Recommend, execute, and monitor multi-step optimization skills \\
    \midrule
    \textbf{Total}   & \textbf{36} & \\
    \bottomrule
  \end{tabular}
\end{table}

All 36 tools share three implementation patterns that promote reliability and LLM self-correction.  First, every parameter carries a typed JSON Schema object with a plain-English description; the \mcp{} server validates incoming calls before executing any handler code.  Second, all tools return a uniform JSON envelope with explicit unit annotations to prevent silent unit-mismatch errors.  Third, failed calls include a hint field with actionable recovery guidance---e.g., \emph{``load the circuit first''}---enabling the LLM to self-correct workflow-ordering errors without user intervention~\cite{qin2024toolllm,schick2024toolformer}.

In the following, we present two example user queries to illustrate how the LLM autonomously orchestrates tool usage.
\begin{chatuser}
\noindent\textbf{User:} \textit{Are there any voltage violations on the feeder?}
\end{chatuser}
\begin{chatbot}
\noindent\textbf{\iotllm{}:} The LLM triggers a power flow solve followed by a system-wide bus voltage scan, filtering buses outside the acceptable range ($0.95$--$1.05$~p.u.). It returns a natural-language summary identifying any violations with per-unit values.\\
\textbf{Tools chained:} solve\_power\_flow $\rightarrow$ get\_all\_bus\_voltages
\end{chatbot}
\begin{chatuser}
\noindent\textbf{User:} \textit{Run a 24-hour simulation and show voltages.}
\end{chatuser}
\begin{chatbot}
\noindent\textbf{\iotllm{}:} The LLM chains load-shape assignment, a 24-hour QSTS run, and a voltage profile extraction, identifying the worst-case interval and bus. Results populate the QSTS dashboard automatically.\\[2pt]
\textbf{Tools chained:} create\_loadshape $\rightarrow$ assign\_loadshape $\rightarrow$ run\_qsts $\rightarrow$ get\_qsts\_voltage\_profile
\end{chatbot}
 The nine equipment-management tools (Capacitor, Reactor, and Regulator/Tap categories) enable closed-loop \emph{what-if} analysis: for example, engineers can add or remove shunt banks, adjust regulator taps ($-16$ to $+16$), and re-solve power flow iteratively---transforming \iotllm{} from a read-only assistant into an interactive design environment.

\vspace{-1em}
\subsection{Skills Framework}

\begin{figure}[t]
\centering
\resizebox{\columnwidth}{!}{\input{tikz/fig_skills_framework}}
\caption{Skills Framework architecture showing query flow from natural language through skill orchestration to \mcp{} tool execution.}
\label{fig:skills_arch}
\end{figure}
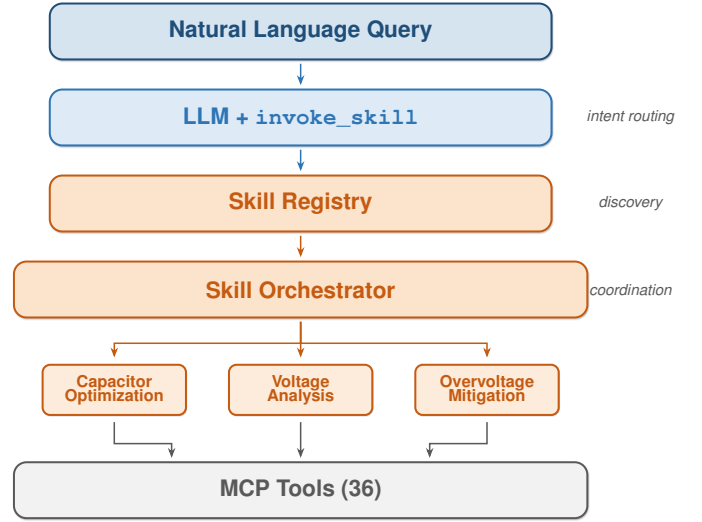

While individual tools are deliberately atomic, practical distribution tasks (diagnosing undervoltage, optimizing capacitor placement, mitigating overvoltage) require coordinating multiple tools in sequence. Delegating this sequencing entirely to the LLM would introduce non-determinism, risking incorrect tool ordering or missed steps in safety-relevant analyses.  To address this, \iotllm{} introduces a \emph{Skills} layer~\cite{zhang2025agentskills}: each skill is a Python class that encapsulates a complete multi-step workflow, calling \mcp{} tools programmatically through an inversion-of-control callback (Fig.~\ref{fig:skills_arch}).  The LLM selects the appropriate skill; the skill itself executes deterministically.

Three skills are currently implemented.  1) The \emph{capacitor placement optimization} skill addresses chronic undervoltage using Particle Swarm Optimization (PSO)~\cite{kennedy1995pso}, proceeding through baseline voltage assessment, candidate bus selection, PSO iteration over (bus, kvar) placement pairs, optimal placement, and post-optimization verification; in preliminary tests on the IEEE 13-bus feeder it substantially reduced undervoltage violations, and Section~\ref{sec:workflow_scale} evaluates it at 123-bus scale.  2) The \emph{voltage violation analysis} skill classifies each bus against standard voltage limits---severe ($>$3\% deviation), moderate (2--3\%), or minor ($<$2\%)---performs root-cause correlation with feeder topology and load density, and returns ranked corrective recommendations.  All bus voltages are reported as positive-sequence magnitudes; per-phase analysis for unbalanced feeders is supported at the tool level but is not currently exposed through this skill.  3) The \emph{overvoltage mitigation} skill resolves buses above 1.05~p.u.~\cite{sun2019review} through a prioritized three-strategy sequence: tap changer adjustment, shunt reactor placement sized by $Q_{\mathrm{reactor}} = (V_{\mathrm{actual}}^{2} - V_{\mathrm{target}}^{2})/X$, and excess capacitor removal.  Preliminary testing resolved all overvoltage buses across the tested scenarios. Table~\ref{tab:tools_vs_skills} summarizes the structural differences between tools and skills.  The key distinction is granularity: tools provide elementary operations from which skills compose deterministic workflows.

\begin{table}[t]
\centering
\caption{Comparison of \mcp{} Tools and \iotllm{} Skills}
\label{tab:tools_vs_skills}
\begin{tabular}{@{}lll@{}}
\toprule
\textbf{Dimension} & \textbf{\mcp{} Tool} & \textbf{Skill} \\
\midrule
Granularity     & Single operation        & Multi-step workflow \\
Algorithm       & None                    & PSO, classification \\
State           & Stateless               & Tracks iterations \\
Calls per use   & 1                       & 5--50+             \\
Determinism     & Deterministic           & Deterministic      \\
LLM role        & Argument extraction     & Skill selection only \\
\bottomrule
\end{tabular}
\end{table}

The tools and skills described above are made accessible through an interactive web platform that unifies conversational control, data access, and integrated visualization, as described next.

%% file: tikz/fig_skills_framework.tex
\begin{tikzpicture}[
  >=Stealth,
  block/.style={
    rectangle, draw, rounded corners=5pt, thick,
    minimum height=0.78cm, minimum width=7.0cm,
    font=\small\bfseries\sffamily, align=center,
    blur shadow={shadow xshift=0.3pt, shadow yshift=-0.5pt,
                 shadow blur steps=4, shadow blur radius=0.8pt}
  },
  skill/.style={
    rectangle, draw, rounded corners=3pt, thick,
    minimum height=0.7cm, minimum width=2.0cm,
    font=\fontsize{6.5}{7.5}\selectfont\bfseries\sffamily, align=center,
    blur shadow={shadow xshift=0.2pt, shadow yshift=-0.3pt,
                 shadow blur steps=3, shadow blur radius=0.5pt}
  },
  arr/.style={-{Stealth[length=4pt,width=3pt]}, semithick, color=arrowcol},
  garr/.style={-{Stealth[length=4pt,width=3pt]}, semithick, color=goldarrow},
  grarr/.style={-{Stealth[length=4pt,width=3pt]}, semithick, color=layer4},
  lbl/.style={
    font=\fontsize{6}{7}\selectfont\sffamily\itshape,
    text=layer4
  },
]

\node[block, fill=layer1lt, draw=layer1, text=layer1] at (0,0) (query) {Natural Language Query};

\node[block, fill=layer2lt, draw=layer2, text=layer2] at (0,-1.2) (llm) {LLM + \texttt{invoke\_skill}};

\node[block, fill=layer3lt, draw=layer3, text=layer3] at (0,-2.4) (registry) {Skill Registry};

\node[block, fill=layer3lt, draw=layer3, text=layer3, minimum width=8.0cm]
  at (0,-3.6) (orch) {Skill Orchestrator};

\node[skill, fill=layer3lt!70, draw=layer3, text=layer3]
  at (-2.6,-5.0) (cap) {Capacitor\\[-2pt]Optimization};
\node[skill, fill=layer3lt!70, draw=layer3, text=layer3]
  at (0,-5.0) (volt) {Voltage\\[-2pt]Analysis};
\node[skill, fill=layer3lt!70, draw=layer3, text=layer3]
  at (2.6,-5.0) (over) {Overvoltage\\[-2pt]Mitigation};

\node[block, fill=layer4lt, draw=layer4, text=layer4, minimum width=8.0cm]
  at (0,-6.4) (tools) {MCP Tools (36)};

\draw[arr] (0,-0.45) -- (0,-0.75);
\draw[arr] (0,-1.65) -- (0,-1.95);

\draw[garr] (0,-2.85) -- (0,-3.15);

\draw[garr] (0,-4.05) -- (0,-4.35) -| (-2.6,-4.55);
\draw[garr] (0,-4.05) -- (0,-4.55);
\draw[garr] (0,-4.05) -- (0,-4.35) -| (2.6,-4.55);

\draw[grarr] (-2.6,-5.45) -- (-2.6,-5.75) -| (-1.8,-5.95);
\draw[grarr] (0,-5.45) -- (0,-5.95);
\draw[grarr] (2.6,-5.45) -- (2.6,-5.75) -| (1.8,-5.95);

\node[lbl] at (4.6,-1.2) {intent routing};
\node[lbl] at (4.6,-2.4) {discovery};
\node[lbl] at (4.6,-3.6) {coordination};

\end{tikzpicture}%

%% file: sections/platform.tex
\section{Platform and Visualization}
\label{sec:platform}
\label{sec:visualization}

\subsection{Architecture and Chat Interface}




The \textbf{UI tier} hosts \textbf{Layer~1} (the Next.js frontend~\cite{nextjs}).
The \textbf{application tier} combines \textbf{Layers~2--3}, namely the LLM Service and the \mcp{} Server, within a single FastAPI container~\cite{fastapi2018}. This tier communicates with the frontend through RESTful APIs and accesses the data tier through provider-specific SDKs and the SQLAlchemy ORM~\cite{sqlalchemy,fowler2003patterns}. The ORM maps Python objects to relational records and generates PostgreSQL queries for managing user sessions, chat histories, circuit metadata, and load profiles. The \textbf{data tier} hosts \textbf{Layer~4}, comprising the \opendss{} simulation engine, PostgreSQL~\cite{postgresql} for relational state, and MinIO~\cite{minio} for S3-compatible circuit-package storage. Docker Compose orchestrates the services, while JWTs authenticate API requests~\cite{rfc7519} and path sanitization is applied to circuit uploads at the API boundary.

\begin{figure}[t]
\centering
\resizebox{\columnwidth}{!}{\input{tikz/fig_web_platform}}
\caption{Three-tier web platform architecture with technology stack.}
\label{fig:web_arch}
\end{figure}
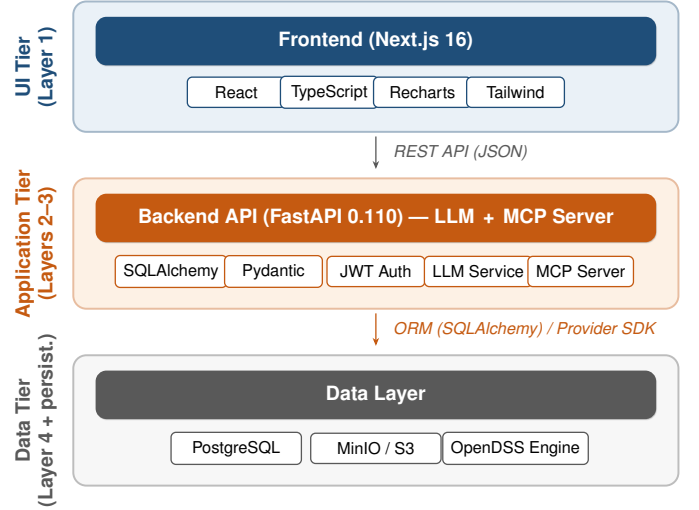

\begin{figure*}[t]
\centering
\includegraphics[width=0.95\textwidth]{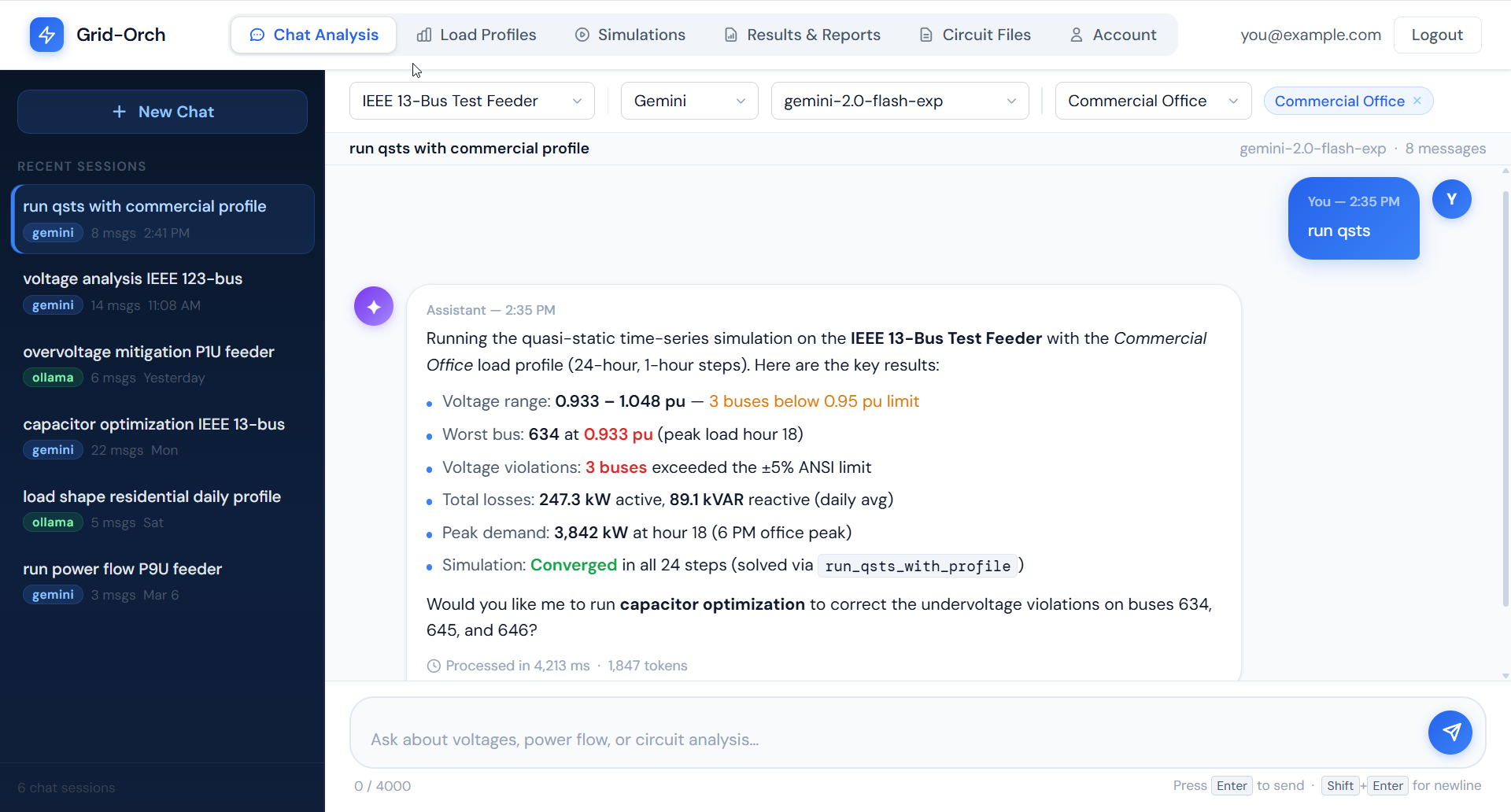}
\caption{\iotllm{} chat interface showing a voltage analysis session. The sidebar lists conversation sessions; the control header provides circuit, provider, model, and profile selectors; the message area displays tool invocations and inline analysis results.}
\label{fig:chat}
\end{figure*}

The chat interface (Fig.~\ref{fig:chat}) provides the primary interaction mode, retaining the familiar structure of general-purpose LLM assistants while integrating power-system-specific controls. A collapsible session sidebar lists the authenticated user’s conversations and, upon selection, restores the complete message history and associated circuit, provider, model, and load-profile context. A compact header provides four inline selectors for configuring the active session’s simulation context. When the LLM invokes one or more \mcp{} tools, a collapsible \emph{Tool Call} panel above the response displays each tool name, its input arguments, and the raw \opendss{} JSON output, enabling engineers to audit the simulation calls supporting the response.

Each tool returns a structured JSON envelope containing a success flag, result data with explicit unit annotations (e.g., per-unit voltages, kW losses), and, on failure, a field with actionable recovery suggestions. The frontend parses these typed JSON responses to automatically select the appropriate chart component. For example, a voltage query triggers a bar chart, while a QSTS result populates the timeseries viewer.

\vspace{-1em}
\subsection{Provider and Data Libraries}

The provider-agnostic LLM layer described in Section~\ref{sec:architecture} supports four model backends, including two locally deployable options for air-gapped operation, so that prompts and simulation data remain within the operator’s network.

Load profiles define the time-varying per-unit scaling applied to loads during QSTS simulations. Ten built-in synthetic profiles derived from the End-Use Load Profiles dataset~\cite{nrel2022eulp} represent residential, commercial-office, data-center, industrial, solar-generation, and worst-case peak-stress scenarios. Engineers may also upload custom profiles as two-column CSV files. The circuit library includes the IEEE 13- and 123-bus test feeders~\cite{schneider2017testfeeders}, together with bus-coordinate data for topology visualization, and supports on-demand retrieval of Smart-DS synthetic feeders~\cite{smartds_dataset}. Each circuit package contains the \opendss{} master DSS file, associated component files, and optional bus coordinates. The package is maintained as a versioned archive, with its metadata registered in PostgreSQL to support simulation reproducibility.

\vspace{-1em}
\subsection{Visualization and User Interface}
\label{sec:viz_topology}

\begin{figure}[t]
  \centering
  \includegraphics[width=\columnwidth]{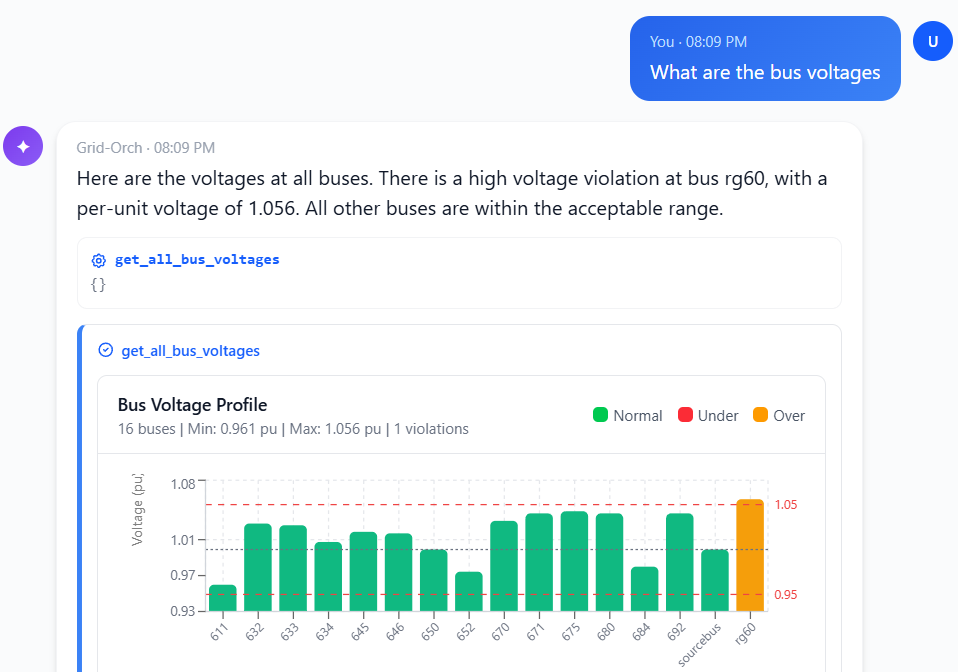}
  \caption{Inline visualization in the \iotllm{} chat interface.  The user's natural-language query triggers a system-wide voltage scan; the response includes a textual analysis and an auto-rendered voltage profile bar chart with per-unit limit shading.}
  \label{fig:chat_voltage}
\end{figure}

\iotllm{} renders simulation results across three integrated surfaces---inline chat, the QSTS dashboard, and exported reports---all sharing a common component library so charts are identical regardless of surface.  Charts are rendered automatically whenever a tool response contains structured data. Fig.~\ref{fig:chat_voltage} illustrates a representative interaction in which the user asks, \emph{``What are the bus voltages?''} \iotllm{} retrieves the system-wide voltage magnitudes and returns both a natural-language summary and an inline voltage-profile bar chart with green, amber, and red bands indicating operating limits. The summary identifies the elevated voltage at bus~rg60, the regulator-output bus, as 1.056~p.u. The collapsible tool-call panel, shown collapsed in the figure, enables engineers to audit every \opendss{} invocation supporting the response.

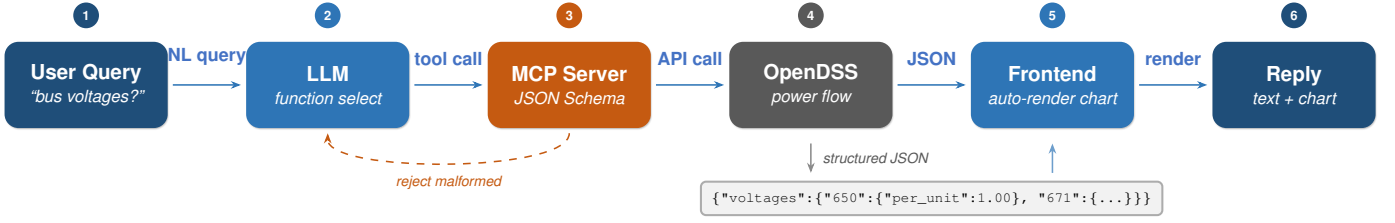
\begin{figure*}[t]
  \centering
  \resizebox{\textwidth}{!}{\input{tikz/fig_e2e_pipeline}}
  \caption{End-to-end pipeline for a natural-language voltage query. The user's prompt traverses six stages: LLM function selection, \mcp{} schema validation, \opendss{} power flow execution, structured JSON return, frontend chart rendering, and grounded textual reply. Malformed tool calls are rejected at step~3 and retried.}
  \label{fig:e2e_pipeline}
\end{figure*}

Fig.~\ref{fig:e2e_pipeline} traces the end-to-end data flow for a representative voltage query through six stages. The user's natural-language prompt is converted by the LLM into a typed \mcp{} tool call (step~2); the \mcp{} server validates the call against its JSON Schema before forwarding to \opendss{} (step~3); \opendss{} solves the power flow and returns a structured JSON result with explicit unit annotations (step~4); the frontend pattern-matches the JSON payload to auto-render the appropriate chart (step~5); and the LLM grounds its textual reply on the same JSON (step~6). If the tool call fails schema validation, the \mcp{} server rejects it and the LLM retries with corrected parameters. Because the reported numerical values are obtained from \opendss{} through schema-validated tool calls, the LLM functions primarily as an interface and interpreter rather than as the source of simulation results, thereby reducing the risk of numerical hallucination. The end-to-end data flow is further summarized in Algorithm \ref{alg:e2e_llm_pipeline}.

\begin{algorithm}[hbt]
\caption{End-to-End LLM--MCP--OpenDSS Interaction Pipeline}
\label{alg:e2e_llm_pipeline}
\begin{algorithmic}[1]
\Require User query $q$ in natural language
\Ensure Grounded textual answer and visualization

\State Receive user query $q$
\State Convert $q$ into an MCP function call with JSON parameters
\State Validate the function call using the MCP server's JSON Schema
\State Execute the validated request in OpenDSS
\State Return structured JSON outputs from OpenDSS
\State Ground the LLM response on the returned JSON outputs
\State Render charts or tables in the frontend from the same typed JSON payload

\end{algorithmic}
\end{algorithm}

\begin{figure}[t]
  \centering
  \includegraphics[width=0.9\columnwidth]{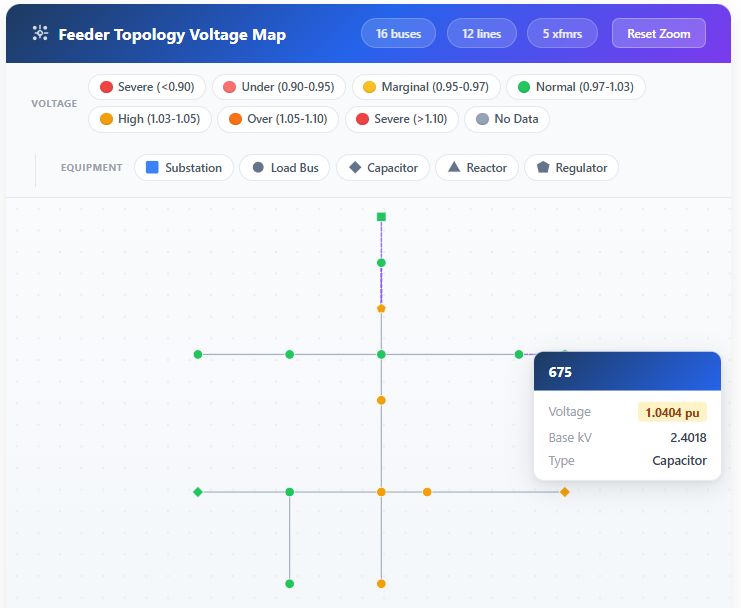}
  \caption{Interactive feeder topology map for the IEEE 13-bus test feeder. Nodes are colored by per-unit voltage with five shapes encoding element types; hovering a node displays its information.}
  \label{fig:topology13}
\end{figure}

The feeder topology map in Fig.~\ref{fig:topology13} renders the IEEE 13-bus feeder as an interactive SVG, using node color to encode voltage and five distinct shapes to distinguish element types. The visualization automatically scales to larger networks, including the IEEE 123-bus feeder (Fig.~\ref{fig:topology}) and the 2{,}356-bus SmartDS feeder.

\begin{figure}[t]
  \centering
  \includegraphics[width=0.49\textwidth]{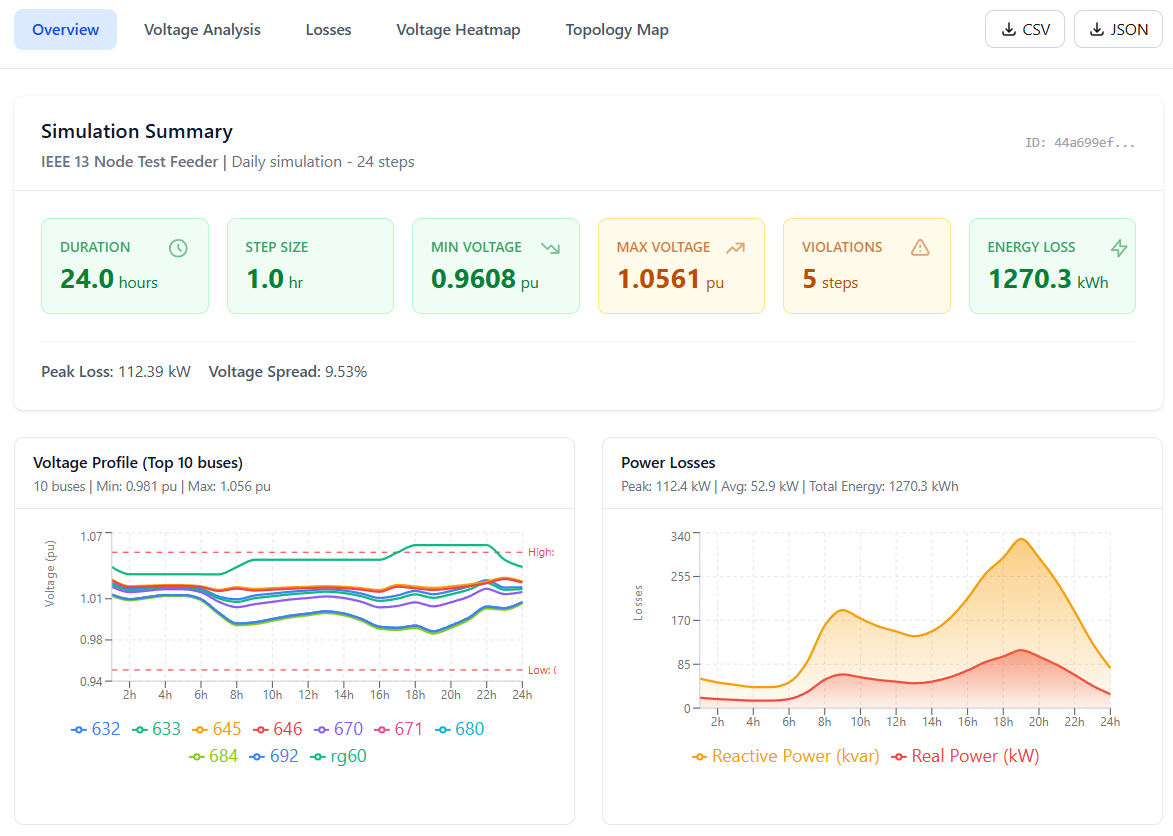}
  \caption{QSTS Dashboard --- Overview tab showing simulation summary KPIs, top-10 bus voltage profile (left), and 24-hour power loss chart (right) for the IEEE 13-bus feeder with residential load pattern.  Five tabs (Overview, Voltage Analysis, Losses, Voltage Heatmap, Topology Map) provide comprehensive post-simulation exploration with CSV/JSON export.}
  \label{fig:qsts_overview}
\end{figure}

\begin{figure}[t]
  \centering
  \includegraphics[width=0.48\textwidth]{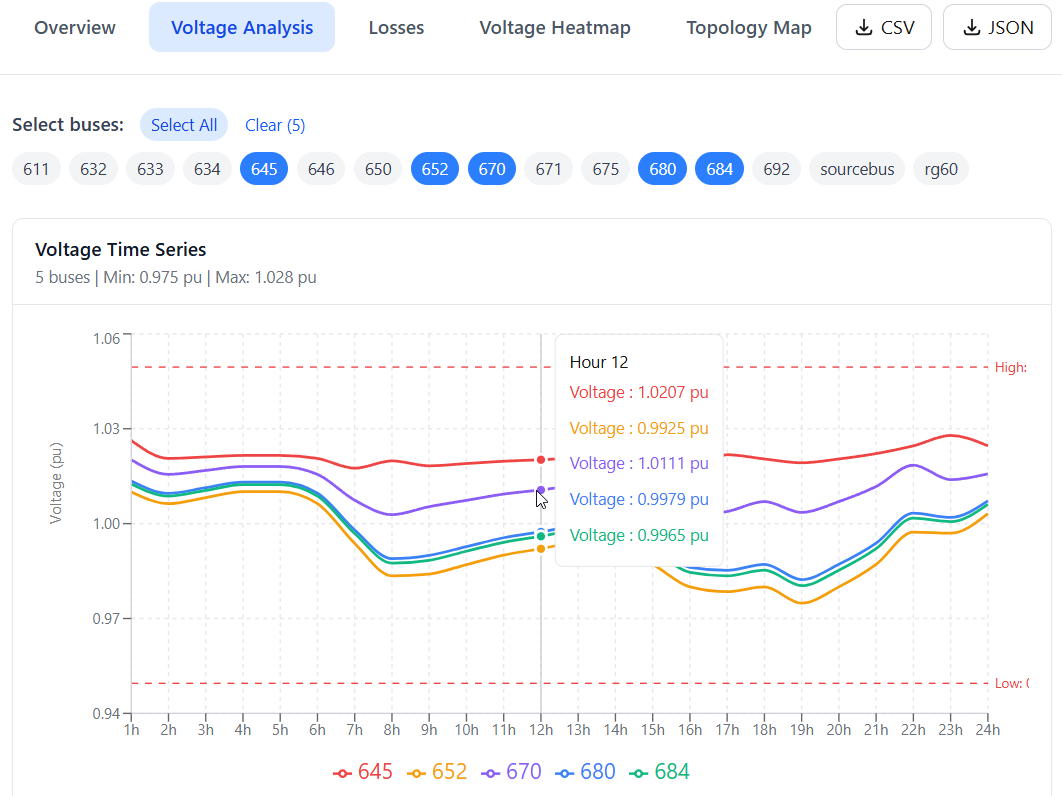}
  \caption{QSTS Dashboard --- Voltage Analysis tab.  Engineers select buses via clickable chips (five buses selected: 645, 652, 670, 680, 684) and inspect their 24-hour voltage timeseries with interactive hover tooltips.  Dashed red lines indicate the upper (1.05~p.u.) and lower (0.95~p.u.) voltage limits.}
  \label{fig:qsts_voltage}
\end{figure}

For post-simulation analysis, a five-tab QSTS dashboard~\cite{montenegro2016qsts}, implemented using the Recharts visualization library~\cite{recharts}, provides a consistent visual experience by reusing the components employed by the inline chat charts (Figs.~\ref{fig:qsts_overview}--\ref{fig:qsts_voltage}). The Overview tab summarizes key performance indicators, including minimum and maximum voltages, the number of voltage violations, and total real- and reactive-power losses. The Voltage Analysis tab enables users to select individual buses through clickable chips and overlays the corresponding time series with interactive tooltips. The Losses tab presents time-varying real- and reactive-power losses as a stacked area chart. The Voltage Heatmap displays a bus-by-time matrix color-coded by per-unit voltage, enabling rapid identification of violation periods. The Topology Map reuses the interactive SVG visualization from the chat interface to depict the spatial distribution of voltage. Each tab supports one-click export of the underlying data in CSV or JSON format. The system persists user sessions, chat histories, and uploaded circuit packages on a per-user basis and supports email/password registration and login through JWT-based authentication~\cite{rfc7519}.

%% file: tikz/fig_web_platform.tex
\begin{tikzpicture}[
  >=Stealth,
  header/.style={
    rectangle, draw, rounded corners=6pt, very thick,
    minimum width=12.0cm, minimum height=1.0cm,
    font=\large\bfseries\sffamily, align=center, text=white,
    blur shadow={shadow xshift=0.4pt, shadow yshift=-0.6pt,
                 shadow blur steps=5, shadow blur radius=1pt}
  },
  comp/.style={
    rectangle, draw, rounded corners=3pt, thick,
    minimum height=0.65cm, minimum width=2.1cm, inner sep=5pt,
    font=\normalsize\sffamily, align=center, fill=white
  },
  tierbg/.style={
    rectangle, draw, very thick, rounded corners=8pt,
    minimum width=13.0cm, inner ysep=8pt, inner xsep=8pt,
  },
  tierlbl/.style={
    font=\large\sffamily\bfseries,
    rotate=90, align=center
  },
  proto/.style={
    font=\normalsize\sffamily\itshape,
    text=layer4
  },
  arr/.style={-{Stealth[length=5pt,width=4pt]}, thick, color=layer4},
  garr/.style={-{Stealth[length=5pt,width=4pt]}, thick, color=goldarrow},
]

\node[tierbg, fill=layer1lt, fill opacity=0.5, draw=layer1!40,
      minimum height=2.8cm] at (0,0) (t1bg) {};
\node[tierlbl, text=layer1] at (-7.3,0) {UI Tier\\(Layer 1)};

\node[header, fill=layer1, draw=layer1] at (0,0.55) (fe)
  {Frontend (Next.js 16)};

\node[comp, draw=layer1] at (-3.0,-0.55) {React};
\node[comp, draw=layer1] at (-1.0,-0.55) {TypeScript};
\node[comp, draw=layer1] at (1.0,-0.55) {Recharts};
\node[comp, draw=layer1] at (3.0,-0.55) {Tailwind};

\draw[arr] (0,-1.5) -- (0,-2.2);
\node[proto, anchor=west] at (0.25,-1.85) {REST API (JSON)};

\node[tierbg, fill=layer3lt, fill opacity=0.5, draw=layer3!40,
      minimum height=2.8cm] at (0,-3.8) (t2bg) {};
\node[tierlbl, text=layer3] at (-7.3,-3.8) {Application Tier\\(Layers 2--3)};

\node[header, fill=layer3, draw=layer3] at (0,-3.25) (be)
  {Backend API (FastAPI 0.110) --- LLM \,+\, \mcp{} Server};

\node[comp, draw=layer3] at (-4.4,-4.4) {SQLAlchemy};
\node[comp, draw=layer3] at (-2.2,-4.4) {Pydantic};
\node[comp, draw=layer3] at (0.0,-4.4) {JWT Auth};
\node[comp, draw=layer3] at (2.2,-4.4) {LLM Service};
\node[comp, draw=layer3] at (4.4,-4.4) {MCP Server};

\draw[garr] (0,-5.3) -- (0,-6.0);
\node[proto, text=goldarrow, anchor=west] at (0.25,-5.65) {ORM (SQLAlchemy) / Provider SDK};

\node[tierbg, fill=layer4lt, fill opacity=0.6, draw=layer4!40,
      minimum height=2.8cm] at (0,-7.6) (t3bg) {};
\node[tierlbl, text=layer4] at (-7.3,-7.6) {Data Tier\\(Layer 4 + persist.)};

\node[header, fill=layer4, draw=layer4] at (0,-7.05) (dl) {Data Layer};

\node[comp, draw=layer4, minimum width=2.8cm] at (-3.0,-8.2) {PostgreSQL};
\node[comp, draw=layer4, minimum width=2.8cm] at (0.0,-8.2) {MinIO / S3};
\node[comp, draw=layer4, minimum width=2.8cm] at (3.0,-8.2) {\opendss{} Engine};

\end{tikzpicture}%

%% file: tikz/fig_e2e_pipeline.tex
\begin{tikzpicture}[
  >=Stealth,
  stage/.style={
    rectangle, draw, rounded corners=5pt, thick,
    minimum height=1.15cm, minimum width=2.2cm,
    font=\footnotesize\bfseries\sffamily, align=center, text=white,
    blur shadow={shadow xshift=0.3pt, shadow yshift=-0.5pt,
                 shadow blur steps=4, shadow blur radius=0.8pt}
  },
  sublbl/.style={
    font=\fontsize{5.5}{6.5}\selectfont\sffamily\itshape,
    text=layer4, align=center
  },
  arr/.style={-{Stealth[length=4pt,width=3pt]}, semithick, color=arrowcol},
  darr/.style={-{Stealth[length=4pt,width=3pt]}, semithick, dashed, color=layer3},
  lbl/.style={font=\fontsize{7}{8}\selectfont\sffamily\bfseries, text=protolbl},
  stepnum/.style={
    font=\fontsize{5}{5.5}\selectfont\bfseries\sffamily,
    circle, inner sep=1.5pt, text=white, minimum size=10pt
  },
  jsoncallout/.style={
    rectangle, draw=layer4!50, thick, rounded corners=2pt,
    fill=layer4lt, inner sep=4pt,
    font=\fontsize{5.5}{6.5}\selectfont\ttfamily, align=left
  },
  groundlbl/.style={
    font=\fontsize{6}{7.5}\selectfont\sffamily\itshape\bfseries,
    text=goldarrow, align=center
  },
]

\node[stage, fill=layer1, draw=layer1] at (0,0) (user)
  {User Query\\\fontsize{6.5}{7.5}\selectfont\sffamily\itshape\mdseries ``bus voltages?''};
\node[stage, fill=layer2, draw=layer2] at (3.3,0) (llm)
  {LLM\\\fontsize{6.5}{7.5}\selectfont\sffamily\itshape\mdseries function select};
\node[stage, fill=layer3, draw=layer3] at (6.6,0) (mcp)
  {\mcp{} Server\\\fontsize{6.5}{7.5}\selectfont\sffamily\itshape\mdseries JSON Schema};
\node[stage, fill=layer4, draw=layer4] at (9.9,0) (dss)
  {\opendss{}\\\fontsize{6.5}{7.5}\selectfont\sffamily\itshape\mdseries power flow};
\node[stage, fill=layer2, draw=layer2] at (13.2,0) (fe)
  {Frontend\\\fontsize{6.5}{7.5}\selectfont\sffamily\itshape\mdseries auto-render chart};
\node[stage, fill=layer1, draw=layer1] at (16.5,0) (out)
  {Reply\\\fontsize{6.5}{7.5}\selectfont\sffamily\itshape\mdseries text + chart};

\draw[arr] ([xshift=2pt]user.east) -- ([xshift=-2pt]llm.west);
\node[lbl, above] at (1.65,0.15) {NL query};

\draw[arr] ([xshift=2pt]llm.east) -- ([xshift=-2pt]mcp.west);
\node[lbl, above] at (4.95,0.15) {tool call};

\draw[arr] ([xshift=2pt]mcp.east) -- ([xshift=-2pt]dss.west);
\node[lbl, above] at (8.25,0.15) {API call};

\draw[arr] ([xshift=2pt]dss.east) -- ([xshift=-2pt]fe.west);
\node[lbl, above] at (11.55,0.15) {JSON};

\draw[arr] ([xshift=2pt]fe.east) -- ([xshift=-2pt]out.west);
\node[lbl, above] at (14.85,0.15) {render};

\draw[darr] ([yshift=-2pt]mcp.south) .. controls +(0,-0.55) and +(0,-0.55) .. ([yshift=-2pt]llm.south)
  node[sublbl, text=layer3, below, midway, yshift=-1pt] {reject malformed};

\node[jsoncallout, font=\fontsize{5.5}{6.5}\selectfont\ttfamily] at (11.55,-1.55) (jsonbox)
  {\{"voltages":\{"650":\{"per\_unit":1.00\}, "671":\{...\}\}\}};

\draw[arr, color=layer4!70, semithick]
  ([yshift=-6pt]dss.south) -- (9.9,-1.0)
  node[sublbl, text=layer4, right, font=\fontsize{5.5}{6}\selectfont\sffamily\itshape, xshift=1pt]
    {structured JSON} -- ([yshift=3pt]jsonbox.north -| 9.9,0);

\draw[arr, color=layer2!70, semithick]
  ([yshift=3pt]jsonbox.north -| 13.2,0) -- ([yshift=-6pt]fe.south);

\node[stepnum, fill=layer1] at (0,0.95)     {1};
\node[stepnum, fill=layer2] at (3.3,0.95)   {2};
\node[stepnum, fill=layer3] at (6.6,0.95)   {3};
\node[stepnum, fill=layer4] at (9.9,0.95)   {4};
\node[stepnum, fill=layer2] at (13.2,0.95)  {5};
\node[stepnum, fill=layer1] at (16.5,0.95)  {6};

\end{tikzpicture}%

%% file: sections/workflows.tex
\vspace{-1em}
\section{Workflow Demonstrations}
\label{sec:workflows}

To illustrate the proposed autonomous workflow, three representative use cases are presented below.

\vspace{-1em}
\subsection{Session Initialization}


At the start of each session, the user selects a feeder from the circuit library, which includes the IEEE 13-bus, 123-bus, or a SmartDS feeder. The corresponding circuit package is retrieved from either local or MinIO cloud storage, and its bus coordinates are registered for topology visualization. During initialization, the LoadShape tools assign a default residential load profile to every load object, making the circuit immediately available for power-flow or QSTS simulation without additional user configuration. Users can subsequently override the default profile, upload a custom CSV file, or modify individual load assignments through natural-language commands.

\vspace{-1em}
\subsection{Use Case 1: DER Interconnection Screening}
\label{sec:workflow_der}

\textbf{Scenario.} A distribution planner needs to evaluate the voltage impact of interconnecting a
2\,MW photovoltaic installation at bus~675 of the IEEE 13-bus feeder. This is a routine but
time-consuming interconnection screening task encountered regularly in distribution planning
departments.

As illustrated in Fig.~\ref{fig:workflow_comparison}(a), the conventional approach requires five sequential manual steps taking several hours. Steady-state voltage compliance is evaluated against standard voltage limits; full IEEE~1547 screening~\cite{ieee2018std1547} (including fault current and protection coordination) is outside the current scope. With \iotllm{}, the engineer submits a single query:

\begin{chatuser}
    {\noindent\textbf{User:} \textit{What is the impact of adding 2\,MW solar PV at bus 675?}}
\end{chatuser}

\iotllm{} autonomously modifies the circuit model, runs a power flow solve, retrieves system-wide bus voltages, invokes the voltage violation analysis skill, and renders a voltage profile chart inline---completing the full sequence in under two minutes. The \mcp{} protocol introduces no numerical distortion: all computation is performed by the \opendss{} engine; the LLM layer handles only tool selection and result interpretation.

\vspace{-1em}
\subsection{Use Case 2: 24-Hour QSTS Analysis}
\label{sec:workflow_qsts}
\begin{chatuser}
    {\noindent\textbf{User:} \textit{Run a 24-hour simulation with residential profile on load 671 and find any voltage violations.}}
\end{chatuser}

\iotllm{} executes a multi-step sequence without further user input: load-shape creation, assignment to load objects, QSTS execution, voltage profile extraction, violation analysis, and inline chart generation.  The QSTS dashboard (Figs.~\ref{fig:qsts_overview}--\ref{fig:qsts_voltage}) is automatically populated: the Overview tab reports 5 violation steps with minimum voltage 0.9608~p.u., while the Voltage Analysis tab lets engineers drill into individual bus timeseries with interactive tooltips.

\vspace{-1.5em}
\subsection{Use Case 3: Automated Voltage Optimization}
\label{sec:workflow_opt}
\begin{chatuser}
    {\noindent\textbf{User:} \textit{Fix the voltage violations using capacitor optimization.}}
\end{chatuser}

\iotllm{} routes the request to the capacitor-optimization skill, which employs PSO to determine optimal capacitor placements while excluding overvoltage buses from the candidate locations. Following the optimal capacitor placement, all nodes in Fig.~\ref{fig:topology13} are rendered in green or yellow, indicating that their voltages remain within the acceptable operating range of 0.95--1.05~p.u. Finally, the LLM synthesizes the results into a plain-language recommendation accompanied by a before-and-after voltage-profile chart.


\vspace{-1.5em}
\subsection{Scalability: IEEE 123-Bus Evaluation}
\label{sec:workflow_scale}

To evaluate scalability, we repeated the workflow suite on the IEEE 123-bus feeder using the same platform configuration, tool library, skills, and prompts. The IEEE 123-bus test feeder follows the standard benchmark settings used in our previous distribution-system studies~\cite{liu2023dlmp,wang2021trilevel}.

\begin{figure}[t]
  \centering
  \includegraphics[width=0.9\columnwidth]{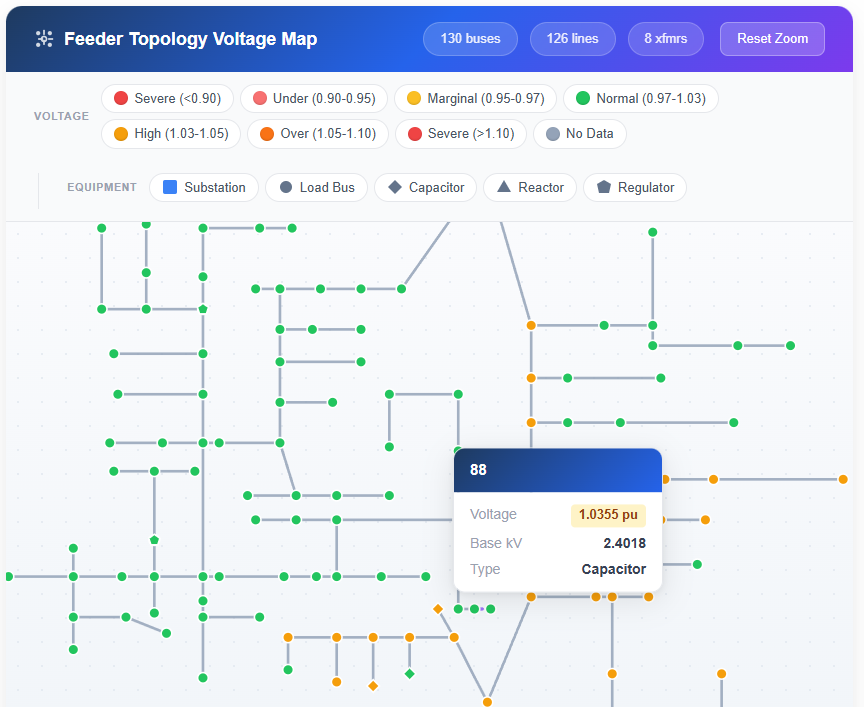}
  \caption{Interactive feeder topology map for the IEEE 123-bus test feeder.}
  \label{fig:topology}
\end{figure}

Table~\ref{tab:scalability} summarizes the measured results.


\begin{table}[!t]
  \caption{Workflow Performance on the IEEE 123-Bus Feeder
  (Cloud: Gemini~2.5 Flash; Local: Llama~3.1~8B via Ollama).}
  \label{tab:scalability}
  \centering
  \renewcommand{\arraystretch}{1.25}
  \setlength{\tabcolsep}{3pt}
  \footnotesize
  \begin{tabular}{p{2.15cm}|p{3.75cm}|c|c}
    \toprule
    \multirow{2}{*}{\textbf{Workflow}} &
    \multirow{2}{*}{\textbf{Result}} &
    \multicolumn{2}{c}{\textbf{Execution Time (s)}} \\
    \cmidrule(lr){3-4}
    & & \textbf{Cloud} & \textbf{Local} \\
    \midrule

    System-wide voltage analysis &
      Minimum, maximum, and mean per-unit voltages match direct \opendss{} scripting. &
      4.2 & 8.3 \\

    24-hour QSTS analysis &
      The complete tool sequence is executed
      autonomously. &
      9.1 & 12.1 \\

    Voltage-violation detection &
      All 28 undervoltage buses are correctly identified, with
      bus-level voltages identical to direct-scripting results. &
      5.3 & 9.5 \\

    Capacitor optimization &
      Resolves all the voltage violations. &
      7.5 & 15.6 \\

    \bottomrule
  \end{tabular}
\end{table}

The complete contingency-analysis session—from feeder selection through violation detection and optimization—was completed in 19.6~s end-to-end, with the latency dominated by LLM inference rather than power-flow computations. The recommended action eliminated all 28 undervoltage violations, demonstrating tool-grounded response generation at a larger system scale. The local model also completed the workflow but required more time for multistep skill invocation on the larger feeder, as reported in Table~\ref{tab:scalability}. The resulting post-optimization feeder topology is shown in Fig.~\ref{fig:topology}.

%% file: sections/discussion.tex
\vspace{-1em}
\section{Discussion and Conclusion} \label{sec:discussion} 

\vspace{-.5em}
\subsection{Discussion and Practical Implications} The results demonstrate that \iotllm{} can coordinate a broad range of distribution-system studies through natural-language instructions while preserving the numerical fidelity of the underlying simulation engine. The demonstrated workflows include system-wide voltage assessment, DER interconnection screening, QSTS analysis, violation detection, and capacitor optimization. In the validation cases, the complete workflows were executed in less than two minutes, and the reported bus voltages, system-level statistics, and identified violations matched those obtained using direct \opendss{} scripting. These results indicate that the proposed agentic interface can reduce the scripting and workflow-integration burden without replacing established power-system simulation methods. 

The numerical agreement can be understood from the separation between language-based orchestration and physics-based computation in the proposed architecture. The LLM does not independently estimate bus voltages, solve the power-flow equations, or generate synthetic simulation results. Instead, it interprets the user's intent, selects tools with predefined input schemas, and organizes their execution. The selected tools then invoke the same \opendss{} models and solution routines used in the corresponding direct scripts. Consequently, provided that the correct circuit, parameters, and analysis sequence are selected, both interaction paths execute the same underlying numerical calculations and should return the same physical quantities. For practitioners, this distinction is important: \iotllm{} introduces a conversational workflow and decision-support layer around a trusted engineering solver rather than substituting an LLM for the solver itself. The resulting outputs therefore remain traceable to the original network model and can be independently verified using conventional \opendss{} scripts. 

In summary, \iotllm{} provides an end-to-end implementation centered on distribution-system analysis, with native in-process optimization skills, an integrated end-user web platform, interactive QSTS visualization, and fully local inference as an explicit design objective for security-sensitive and air-gapped environments. Thus, the principal contribution of \iotllm{} is not only simulator connectivity, but also the integration of natural-language interaction, deterministic engineering workflows, optimization, visualization, and local deployment within a unified operational framework. 

\vspace{-1em}
\subsection{Limitations and Future Work} Despite the demonstrated performance, LLM reliability remains an important consideration for engineering applications. In multi-tool workflows, a model may select a semantically related but incorrect tool, misinterpret an intermediate result, or infer parameter values that were not specified by the user. \iotllm{} mitigates these risks at three levels. First, the \mcp{} boundary restricts the model to 36 explicitly defined tools with structured input schemas and does not expose unrestricted code execution. Second, the Skills framework encapsulates common multi-step studies in predefined execution procedures once the appropriate skill and parameters have been selected. Third, structured error messages provide actionable feedback that allows the agent to revise invalid tool calls without immediate user intervention. Nevertheless, these mechanisms reduce rather than eliminate LLM-related errors. The generated studies should therefore be regarded as engineering analyses that require review, particularly before they are used to support investment decisions, protection-setting changes, or real-time operational actions. 


Scalability of the tool is another open issue. As the tool catalog expands, overlapping tool descriptions and similar parameter schemas may reduce selection accuracy. Long workflows may also accumulate errors across multiple reasoning, tool-execution, and result-interpretation steps, eventually exceeding the model's ability to recover automatically. In addition, the present demonstrations focus primarily on distribution-system steady-state and QSTS applications; they do not yet establish reliability across the broader range of dynamic, protection, and multi-simulator studies encountered in utility practice. Future research will pursue three main directions. First, the tool and skill libraries will be extended to transient-stability, electromagnetic-transient, and protection-coordination studies, with additional simulation engines---including ANDES~\cite{cui2020hybrid}, CYME, PSCAD, GridLAB-D, and PSS/E---connected through the \mcp{} server interface. Selected PowerAgent tools~\cite{zhang2025poweragent} will also be investigated as interoperable components within the \iotllm{} orchestration framework. Second, formal user studies with utility engineers and system analysts will be conducted to quantify usability, task-completion time, error rates, and the level of engineering expertise required to operate the platform. Third, a semantic tool-retrieval layer will be developed to expose only the tools relevant to each query, rather than presenting the complete catalog to the LLM. This approach is expected to improve tool-selection accuracy, reduce prompt complexity, and support expansion to substantially larger multi-simulator tool libraries. 

\vspace{-1.2em}
\subsection{Conclusion} 
\vspace{-0.5em}
This paper presented \iotllm{}, an \mcp{}-based agentic platform for conversational distribution-system analysis and optimization. The platform integrates 36 \opendss{} tools spanning eleven functional categories, three optimization skills, a multi-provider LLM backend with fully local and air-gapped deployment options, and a web-based user environment with inline visualization and an interactive QSTS dashboard. The case studies demonstrated that natural-language requests can be translated into complete engineering workflows, ranging from DER interconnection screening to QSTS violation detection and corrective optimization. For the reported validation cases, the resulting numerical outputs matched those obtained through direct \opendss{} scripting because both approaches relied on the same network models and solver routines. 

Overall, \iotllm{} demonstrates a practical architecture in which an LLM serves as an orchestration and interaction layer while established power-system software remains responsible for the underlying physical calculations. By reducing dependence on custom scripting, integrating multi-step analyses into reusable skills, and supporting fully local deployment, the proposed \iotllm{} platform lowers the barrier to advanced distribution-system studies while retaining solver-level traceability and human engineering oversight. These capabilities provide a foundation for secure, extensible, and practitioner-oriented agentic tools for future power-system planning and operation.